\RequirePackage{fix-cm}
\documentclass{svjour3}                     
\smartqed  
\usepackage{graphicx}
\usepackage{ulem}
\usepackage{amsfonts}
\usepackage{amsmath}
\usepackage{amssymb}
\usepackage{bm}
\usepackage{color}
\usepackage{ctable}
\usepackage[multiple]{footmisc}
\usepackage{mathrsfs}
\usepackage{pdflscape}
\usepackage{rotating}
\usepackage{scalerel}
\usepackage{tabulary}
\usepackage{hyperref}
\hypersetup{
    colorlinks=true,
    citecolor=blue,
    linkcolor=magenta,
    filecolor=blue,      
    urlcolor=blue,
}

\journalname{General Relativity and Gravitation}
\begin{document}

\title{\large Editorial Note to:\thanks{This work is part of a project that
    has received funding from the European Research Council (ERC)
    under the European Union's Horizon 2020 research and innovation
    programme (grant agreement ERC advanced grant 740021--ARTHUS, PI:
    TB). \\ TM is supported by the FONDECYT Iniciaci\'on grant No.11190854.} \\ 
    On the Newtonian Limit of Einstein's Theory of Gravitation\\
    (by J\"urgen Ehlers)}

\titlerunning{Editorial Note to: On the Newtonian Limit of Einstein's Theory of Gravitation} 

\author{Thomas Buchert \and Thomas M\"adler}
\authorrunning{T. Buchert and T. M\"adler} 

\institute{\\ Thomas Buchert \\ \email{buchert@ens-lyon.fr}
\at
Univ Lyon, Ens de Lyon, Univ Lyon1, CNRS, Centre de Recherche Astrophysique de Lyon UMR5574, F--69007, Lyon, France \\ \linebreak
Thomas M\"adler \\ \email{thomas.maedler@mail.udp.cl}
\at          
Escuela de Ingenier\'ia en Obras Civiles \& N\'ucleo de Astronom\'ia, Facultad de Ingenier\'ia y Ciencias, Universidad Diego Portales, Av. Ej\'ercito 441, Santiago, Chile
}

\date{Received: date / Accepted: date}

\maketitle

\begin{abstract}
We give an overview of literature related to J\"urgen Ehlers' pioneering 1981 paper on Frame theory---a theoretical framework for the unification of General Relativity and the equations of classical Newtonian gravitation. 
This unification encompasses the convergence of one-parametric families of four-dimensional solutions of Einstein's equations of General Relativity to a solution of equations of a Newtonian theory if the inverse of a causality constant goes to zero. 
As such the corresponding light cones open up and become space-like hypersurfaces of constant absolute time on which  Newtonian solutions are found as a limit of the Einsteinian ones. 
It is explained what it means to not consider the `standard-textbook' Newtonian theory of gravitation as a complete theory unlike Einstein's theory of gravitation. 
In fact, Ehlers' Frame theory brings to light a modern viewpoint in which  the  `standard' equations of a self-gravitating Newtonian fluid are Maxwell-type equations.
The consequences of Frame theory are presented for Newtonian cosmological dust matter expressed via the spatially projected electric part of the Weyl tensor, and for the formulation of characteristic quasi-Newtonian initial data on the light cone of a Bondi-Sachs metric.
\end{abstract}
\section{The Newtonian Limit}

To set the spirit of J\"urgen's paper \cite{GO}: whenever a notion such as ``Newtonian limit of General Relativity'' appears to be used as if it were standard, trivial or just textbook knowledge, J\"urgen's fine sense of physics immediately raised the important question of whether this is indeed so, 
leaving wide open the possibility of understanding something else under this seemingly common sense expression. This happened often in personal interaction of one of us (TB\footnote{This interaction 
essentially began in 1984 at the Max-Planck-Institut for Astrophysics in Garching, where J\"urgen supervised my (TB's) master and PhD work, and later resulted in several joint papers, among them also papers related to the present subject \cite{buchertehlers:lagrange,buchertehlers,ehlersbuchert:lagrange,ehlersbuchert:weyl}.}) with J\"urgen and it is this way of questioning that keeps the sense of physics awake.  To give another example: in some work I (hereafter, I refers to TB) had hidden assumptions in what I called ``relativistic ideal gas''. J\"urgen did not just use the word ``wrong'', but surprisingly raised the question of what this expression means to me, developing from scratch on the blackboard elements that should define an ideal gas, as if this were his first thoughts of curiosity about this notion.

The paper \cite{GO} is such an example of thoroughly asking this question. 
This pioneering  article laid the ground for fruitful follow-up developments in rendering the notion of ``Newtonian limit'' precise.
Indeed, J\"urgen calls practical implementations of the Newtonian limit empirical (or heuristic), which is ``useful for preliminary considerations [...] while making a suitable ansatz for the metric and matter variables in which small quantities are neglected [...] i.e. they are set to zero.\cite[p. 2]{GO}''.
He emphasizes in his introduction to the problem the absence of a corresponding mathematical justification in the literature. He spells out that empirical tests of General Relativity already presuppose the existence of a limiting process.

Although later publications provide a shortcut or incorporate new developments of what is known as ``Ehlers' Frame Theory'', i.e., a theory that comprises both a
Newtonian ``theory'' and General Relativity within one framework,\footnote{The reader is directed, e.g., to J\"urgen's later papers \cite{ehlers:newtonianlimit,ehlers:examples}, the summary of Todd Oliynyk and Bernd Schmidt \cite{OliynykSchmidt}), or the compact paragraph in the paper \cite{ehlersbuchert:weyl}.} the present translation of his 1981 paper \cite{GO} reveals best J\"urgen's thoughts when generalizing the work by \'Elie Cartan, Kurt Friedrichs, and others.\footnote{For references see section 1 
in J\"urgen's paper \cite{GO}.}
Therein,  it allows the reader to enter his strategy of exploring and discovering afresh a notion that is thought---within a broad community that works empirically with the Newtonian limit---to not contain any ambiguous issue.

The reason why again the Newtonian ``theory'' is put into quotations above reflects J\"urgen's constant reminders that the equations of Newtonian gravitation in their continuous formulation do not represent a theory.\footnote{Rather, we may call it an ``incomplete theory'' for the reasons listed below. J\"urgen thought of the hyperbolic character of a theory that completely determines the system from initial data only. We may add the property of Einstein's theory being \textit{background-free}, in contrast to Newtonian gravitation. (As the constraint equations of Einstein's theory require the specification of boundary conditions (in addition to the initial conditions) the global topology is to be specified too, see the discussion in \cite{Levin}. Einstein's theory in its classical formulation determines the global topology for all times by the topology of the initial Cauchy hypersurface; generally, however, one does not consider the possibility of a dynamical topology change.)} 
There are elements that leave solutions to these equations non-unique, so that it makes no
physical sense even to write down the Euler-Poisson system of equations without specifying and using three elements: 
(i) the introduction of a reference background space (see, however, \cite{Frank} for the case of vacuum where \textit{a priori} no background is present), 
(ii) boundary or fall-off conditions on the deviations from the reference background that need to be specified at all times, and (iii) translation invariance of Newton's equations. 
A thorough investigation of these three elements that make Newton's equations a theory can be found in \cite{buchertehlers}. 
This latter paper was largely motivated from J\"urgen's side by rendering the architecture of Newtonian cosmologies precise, guided by the uniqueness problem of Newton's equations in continuum form, from my side motivated by the averaging problem in cosmology. 
This is also reflected by the {\it degeneracy} of the Newtonian limit, summarized in compact form in the language of fibre bundles and the contraction of the Lorentz group to the Galilei group, that J\"urgen explains in the introduction. We will come back to this later.

Considering the Newtonian limit {\it in terms of equations and geometrical concepts} lies behind Frame theory, i.e., using the same geometrical language for both as the starting point, and assigning the parameter $\lambda = c^{-2} > 0$ to General Relativity and $\lambda =0$ to the limiting theory. 
In other words, Frame theory renders the Newton-Cartan limit of General Relativity mathematically precise. 
J\"urgen's approach to Frame theory is in fact a special (and less formal) application of a more general procedure \cite{geroch} to consider the limit of space-times.
As J\"urgen also points out, Frame theory may help in improving perturbation schemes incorporating relativistic corrections---and indeed his work served later to improve on post-Newtonian expansions \cite{OliynykSchmidt}.
It already  surfaces in  \cite{GO} whose  essential aim is to understand the limit of a family of solutions, and not only the limit of the physical laws. 
Frame theory allows us to explore both. 
As we know, some Newtonian solutions have no relativistic counterpart,
and some general-relativistic space-times have no Newtonian solution as limit. 
J\"urgen enriched this step by focussing on such solutions in his 1997 paper \cite{ehlers:examples}. 
But, already at the end of \cite[Sec.~4]{GO},  we learn about
solutions, e.g., that spherically symmetric isolated relativistic
stars have a corresponding Newtonian star model as limit, and that Schwarzschild as well as Kerr black holes degenerate in the Newtonian limit to the field of a point mass, and J\"urgen discusses the limit of a plane gravitational wave. Frame theory also led to an existence theorem for a class of stationary axisymmetric solutions of General Relativity \cite{Heilig}. 
J\"urgen emphasizes, however, that the interest in forming a notion of the limit through examples should rather focus on non-stationary situations with radiation. 
The first investigations in this regard were made by Winicour, soon after publication of \cite{GO}, who discussed the construction of initial modes based on a Newtonian model for the radiative Bondi-Sachs metric of General Relativity. 
The basic idea behind this construction is outlined in
Section~\ref{sec:5th}, and some consequences of such an initial model with regard to the (radiative) asymptotic structure of General Relativity are briefly mentioned in Section~\ref{sec:close}.

\section{Two Theories}

People have often pointed out that the concepts behind General Relativity and Newtonian physics are entirely different, and
one has to understand not only formally how the Newtonian equations arise through a limiting process from the geometrical theory of Einstein, but also how the content of concepts translates from one to the other. 
Here we do not necessarily enter such philosophical terrain with this question. J\"urgen mentions related thoughts by Kuhn and Feyerabend, who suggest that the existence of a limiting process is not obvious: Einstein's theory rather replaces Newton's for the description of gravitation as the common physical object. 
It is true that Frame theory already uses concepts of General Relativity and formulates Newton's laws accordingly. 
J\"urgen argues in \cite{GO} that with the transition to a formally
simpler theory implications of the more complicated theory get lost;
``the {\it formal} simplification is connected with a {\it conceptual}
complication \cite[p.~16]{GO}''. But, he also stresses that with the
help of Frame theory the meanings of basic notions of
the two theories can be compared, not only the formal structures,
``despite Kuhn and Feyerabend \cite[p.~16]{GO}''.
J\"urgen carefully collects common concepts and assigns them to the, within the Frame theory, common geometrical object of {\it connection}. He emphasizes that both theories are based on a symmetric linear connection for those aspects of the motion of matter that correspond to the notions of inertia or ponderousness (gravity), thereby revealing the common foundation of equivalence of inertial and ponderable (gravitational) mass.\\

I  remember that J\"urgen nuanced his view in discussions of the Newtonian limit later: ``we should either work with General Relativity {\it or} with Newtonian laws to describe a self-gravitating system'', which sounded as  if there remain
incommensurabilities of at least some conceptual aspects of a ``limiting procedure''. It is also a matter of physical interpretation. For example, both on the level of equations and solutions, the same equations and solutions govern an expanding homogeneous-isotropic dust fluid in both theories, but in General Relativity it is interpreted as an expanding space within a Lorentzian space-time, while in Newtonian cosmology it is a fluid expanding within an embedding Euclidean vector space. Although Frame theory delivers an unambiguous answer, mathematically, his statement may reflect aspects of the degeneracy of the limit, which we are going to discuss now.

\section{Opening-up light cones---part I}

Another element that is nowadays crucial for any theory is causality. 
That said, the Newtonian ``limit'' appears as a step back: why should we at all construct the limit where, in J\"urgen's words ``the light cones open up \cite[p.~3]{GO}", thus sending the causal propagation speed to infinity and losing another essential element of a ``theory''.
J\"urgen and I (TB) once discussed in private the ideas of gravitoelectromagnetism (see, e.g., \cite{jantzen}), 
and came to the observation that the first steps towards a finite
propagation speed of gravitation may have been made by Heaviside
\cite{heaviside} around the time of his derivations of key-stones of
classical electrodynamics (such as his derivation of the
Lorentz force). Especially, Heaviside's idea of a gravitational analogy to electrodynamics pointed to a feature of 
Frame theory---Equations (15), (18) and (19) in \cite{GO}---by suggesting a Maxwellian form of the gravitational equations in the limit of setting the parameter $\lambda = c^{-2}$ of the Frame theory to zero.
J\"urgen points out the existence of terms missing in standard writings of Newton's equations, featuring a vector field $\boldsymbol{\omega}$, where the limiting theory assumes the classical Newtonian form iff
$\boldsymbol{\omega}=0$.\footnote{As an aside we may acknowledge that the set of equations (18) and  (19) in \cite{GO}
suggest the presence of a non-inertial repulsive term $\omega^2$ in the field equation, where a cosmologist of the 
``dark energy era'' would be tempted to analyze a possible replacement of a cosmological term, and it suggests that the force field is non-conservative featuring a source for the curl of the gravitational field strength.} 
It has to be emphasized that J\"urgen's field $\boldsymbol{\omega}$ in Equations (15), (18) and (19) results from 
non-inertial force terms
which have to be added to the classical Newtonian system commonly written in non-rotating coordinates. 
The presence of these additional terms implies the non-existence of global inertial systems, which is also a property of Einstein's theory. In J\"urgen's words the limiting theory for 
$\lambda =0$ reduces to Newton's equations, 
``...if the harmonic axial vector field $\boldsymbol{\omega}$, i.e.\ the vorticity field, is constant in every hypersurface $t=const$, so that $\boldsymbol{\omega}$ depends only on $t$.
That is to say, only in this case is there a subset of {\it non-rotating coordinate systems} in the class of Galilean coordinate systems, with respect to which $\boldsymbol{\omega}=0$, ...\cite[p. 8]{GO}''.
However, since the field $\boldsymbol{\omega}$ is a harmonic vector field, this is exactly the freedom that is fixed by the three elements mentioned above that render the Newtonian equations a theory \cite{buchertehlers}. J\"urgen would have certainly added a remark after the results of the above paper that identifies the harmonic freedom in Newton's equations. The harmonic part in the gravitational potential is for example fixed to be zero in the following case \cite{buchertehlers}: introducing a reference background,
the solutions to Poisson's equation for the deviations off this background are unique, 
$\Phi = \Phi_b + \phi$, $\phi' = \phi + Z$, 
with the potential $Z$ obeying the Laplace equation, iff the integral
over the source $\delta\varrho : = \varrho - \varrho_b$
vanishes. Imposing periodic boundary conditions on the deviation
fields, i.e., assuming a $3-$torus topology for the deviation fields,
leaves only spatially constant solutions of the Laplace
equation. Translation invariance of Newton's equations then allows
one to set, without loss of generality, the harmonic potential to zero. A similar argument for a
rigidly rotating $3-$torus holds for the harmonic vector field $\boldsymbol{\omega}$. In General Relativity, exact statements on the harmonic parts are more involved and can be understood through Hodge-de Rham theory \cite{rza4}.

We finally quote J\"urgen again: when formulating both theories geometrically, 
``the ``only" difference in the formulation of the two theories is that the dimensional measures of space-time in Newton's theory are given by a so-called Galilei Structure [...], while in Einstein's Theory they are described by a Lorentz metric.\cite[p. 2]{GO}", i.e., a Lorentz-covariant Newtonian theory would alter the big picture essentially.
But, is this really the whole story, and would this remove the degeneracy of the Newtonian limit? 
We will see below that the answer is not in the affirmative; this
statement concerned the formal structure of the equations.
There is a subtle difference that is related to the connection, which in Newtonian theory is {\it integrable}. 

Before we discuss this important point, we take up the gravitoelectromagnetic analogy again as a first, and a transformation of the Newtonian equations to Lagrangian coordinates as a second, argument to illustrate that
formal considerations of the Newtonian limit on the basis of equations depends on how we look at Newton's equations, and this gives a higher voltage to J\"urgen's emphasis on considering the limit of {\it solutions}. 
Note in this context the first general result on the existence of a Newtonian limit for a large class of {\it solutions} of the Vlasov-Einstein system \cite{Rendall} and subsequent developments \cite{OliynykSchmidt}.  
What we can learn from the Frame theory is the need for a covariant formulation as exemplified by the Newton-Cartan theory, simply because common considerations of Newton's equations are based on writing the equations in coordinate components.

\section{Newton's theory written in coordinate components}

In order to illustrate the various forms we could expect from writing Newton's equations in coordinate components,
we provide two forms presented in the following excursions. Through the trivial aspect of coordinate dependence we will discover important aspects related to the Newtonian limit, where the second form belongs to a development that 
J\"urgen too followed in depth in the 1990's.

\subsection{Eulerian form of Newton's equations}

The first form is to take up the gravitoelectromagnetic analogy hitherto discussed.
Already Newton's equations written in a non-rotating (globally inertial) coordinate system produce a ``Maxwellian picture'' that goes beyond what is usually stated. 
They hide a gravitomagnetic field that, if put to zero, will result in a highly restricted class of solutions. To see this briefly, we may write the Euler-Poisson system in modernized ``Heaviside language''. We recall
the Euler-Poisson system for a self-gravitating restmass density field $\varrho ({\bf x},t)$, $\varrho \ge 0$, and a velocity field ${\bf v}({\bf x},t)$ (for a dust matter model), represented 
in a non-rotating Eulerian coordinate system ${\bf x}$ in the Newton-Galilei space time and foliated into Euclidean space sections labelled by $t$, evolving out of initial data at time $t=t_{\mathrm{ini}}$, the evolution equations,
\begin{eqnarray}
\label{ENS1}
\frac{\partial}{\partial t}
{\bf v} \;= - ({\bf v} \cdot \boldsymbol{\nabla}) {\bf v} + {\bf b} \;\; ;
\;\;{\bf v}({\bf x}, t_{\mathrm{ini}}) = :{\bf V}_{\mathrm{ini}} \;\;,\\
\frac{\partial}{\partial t}
\varrho \;= - \boldsymbol{\nabla} \cdot(\varrho {\bf v}) \;\; ; \;\;
\varrho({\bf x}, t_{\mathrm{ini}}) = :\varrho_{\mathrm{ini}}  \;\; ,
\label{ENS2}
\end{eqnarray}
and Poisson's equation for the gravitational potential $\Phi$,
\begin{equation}
\label{Poisson}
\Delta \Phi = \Lambda - 4\pi G \varrho \;\;,
\end{equation}
where ${\bf b}({\bf x},t)= {\bf F}/{m_I}$ is the acceleration field related to a force field $\bf F$ with respect to an inertial 
mass $m_I$, and $\Lambda$ is the cosmological constant. The equivalence of inertial ($m_I$) and gravitational ($m_G$) mass asserts that the gravitational field strength
${\bf g}({\bf x},t)= {\bf F^G}/{m_G} = :- \boldsymbol{\nabla}\Phi$ substitutes the acceleration field ${\bf b}$ in Euler's equation \eqref{ENS1}, yielding a closed system.
Thus, the evolution of the self-gravitating continuum is constrained by Newton's
field equations for the gravitational field strength ${\bf g}({\bf x},t)$:
\begin{equation}
\boldsymbol{\nabla} \cdot {\bf g} = \Lambda - 4 \pi G \varrho \quad;\quad
\boldsymbol{\nabla} \times {\bf g} = {\bf 0} \;\;,\label{ENSelectric}
\end{equation} 
which according to textbooks is the point where the analogy with (electrostatic) Maxwell's field equations for the electric field strength stops. 
Going a bit further, however, we find via the evolution equation for the field strength $\bf g$ a 
complementary set of Maxwell-type equations:
we introduce the current density ${\bf j}: = \varrho {\bf v}$ for the flow of fluid elements, and 
we may ask about its relation to the time--derivative
of the gravitational field strength. Using the continuity equation \eqref{ENS2} and 
the relation between the restmass density and the divergence equation for the 
field strength, first equation of \eqref{ENSelectric}, we immediately find (for more details see \cite{buchert:integral}):
\begin{equation}
\frac{\partial}{\partial t}\varrho = - \boldsymbol{\nabla}\cdot {\bf j} \;\;\;{\rm and}
\;\;- 4\pi G \frac{\partial}{\partial t}\varrho = \boldsymbol{\nabla}\cdot 
\frac{\partial}{\partial t}{\bf g}\;\;\;.
\end{equation}
We conclude:
\begin{equation}
\label{g-evolution}
\boldsymbol{\nabla}\cdot [\; \frac{\partial}{\partial t}{\bf g} - 4\pi G {\bf j}\; ]\;=\;
0\;\;\;\Rightarrow\;\;\; 
\frac{\partial}{\partial t}{\bf g} - 4\pi G {\bf j}\;=:\;\boldsymbol{\nabla}
\times\boldsymbol{\tau}\;\;\;.
\end{equation}
The vector field $\boldsymbol{\tau}$ can be interpreted as a gravitomagnetic field strength. Since its divergence is not fixed, we may also assume---in analogy to electromagnetism---that there are no gravitomagnetic monopoles, so that we obtain the following full set in Maxwellian form:
\begin{eqnarray}
\label{generalnewton}
&\boldsymbol{\nabla} \cdot {\bf g} = \Lambda - 4 \pi G \varrho\;\;;\;\;
&\boldsymbol{\nabla} \cdot \boldsymbol{\tau} = 0 \;\;, \nonumber\\
&\boldsymbol{\nabla} \times {\bf g} = -\frac{1}{c^2}\frac{\partial}{\partial t} {\boldsymbol{\tau}} \quad;\quad\,
&\boldsymbol{\nabla} \times \boldsymbol{\tau} = \frac{\partial}{\partial t} {\bf g}
- 4\pi G {\bf j} \;\;.
\end{eqnarray}
In the above set of equations we added {\it ad hoc} to the field equation for $\bf g$ the term corresponding to Heaviside's
``missing term" (the source term for the curl of the field strength $\bf g$ that makes the analogy complete and the speed $c$ of propagation of the gravitational interaction finite,
and which would render the gravitational force non-conservative). Note that this term is nothing but the gravitomagnetic analogy to Maxwell's displacement current, on which Heaviside's argument is based.
We notice that, by dropping again this term (equivalent to sending $c$ to infinity), Newton's equations still feature a 
gravito\-magnetic field.
Setting this field to zero or restricting it to a harmonic vector field would result in a highly restricted class of solutions to Newton's equations (e.g., for irrotational
flows the gradient of the density has to be aligned with the velocity).
The above set of equations shows that Newton's theory can come with a different face, the gravitomagnetic field 
arises only implicitly, rather than explicitly, in the limit of the Frame theory.  
As for the analogy to Maxwell's displacement current we may speculate that Newton would have included it, had he lived at a later epoch, hence the question of the light cones opening-up would not arise in a formulation of the limiting process of General Relativity to a Lorentz-covariant vector theory of gravitation.\footnote{The reader may consult the rich literature on gravitoelectromagnetism, e.g. \cite{jantzen}, emphasizing that the notion of Newtonian limit also crucially depends on the considered families of space-time splits.}

\subsection{Lagrangian form of Newton's equations}

The second form we discuss now arises by transforming the Euler-Poisson system to Lagrangian coordinates, 
represented by a one-parametric family of spatial diffeomorphisms between the Lagrangian spatial coordinates $ \mathbf{X} $  which label fluid elements, and the Eulerian ones $ \mathbf{x} $, which are now conceived as values of a position field $ \mathbf{f} $ of these elements at Newtonian time $t$:\footnote{Latin indices in the middle of the alphabet $i,j,k,... $ denote coordinate indices $i,j,k,...\in\{1,2,3\}$ for three-dimensional coordinates $X^i$. A vertical slash ${}_{\vert i}$ stands for the partial spatial derivative with respect to Lagrangian coordinates $X^i$, while a comma denotes partial spatial derivative with respect to Eulerian coordinates $x^i$. Antisymmetrization of an index pair is given by $[ij] = (ij-ji)/2$, and $\epsilon_{ijk} $ is the Levi-Civita symbol. Latin bold face indices in the beginning of the alphabet $\mathbf{a,b,c, }...$ label the components of a triad $\mathbf{a,b,c}...\in\{1,2,3\}$. Later, latin indices $a,b,c$ are used as four-dimensional space-time indices $a,b,c...\in\{0,1,2,3\}$, while latin capital indices $A,B,C...$ are two-dimensional, $A,B,C,...\in\{2,3\}$, for angular coordinates of a $2-$sphere. The four-dimensional covariant derivative is denoted with the `nabla' symbol $\nabla$. We use the Einstein summation convention throughout. To be in tune with \cite{GO}, we choose the $4-$metric signature as $(+,-,-,-)$.}
\begin{equation}
\mathbf{f}_t :\; \mathbb{R}^3 \longrightarrow \mathbb{R}^3 \;\;;\;\;
             \mathbf{X} \longmapsto  \mathbf{x}=  \mathbf{f}( \mathbf{X},t) \quad \mbox{and} \quad \mathbf{X} := \mathbf{f} (\mathbf{x}, t_{\mathrm{ini}})\;,
\label{trajectory}
\end{equation}
with the determinant of the Jacobian matrix, $(f^{\mathbf{a}}_{\ \vert j})$, describing the local volume deformation of the dust matter,
\begin{equation}
J := \det \left( f^{\mathbf{a}}_{\  \vert j} (\mathbf{X},t)\right) = \frac16 \epsilon_{\mathbf{abc}}  \epsilon^{ijk} f^\mathbf{a}_{\ \vert i}f^\mathbf{b}_{ \ \vert j} f^\mathbf{c}_{\ \vert k}\:\:,\;\;J(\mathbf{X}, t_{\mathrm{ini}}):=1\;\;.
\end{equation} 
To obtain the Lagrangian formulation of Newtonian gravitation, we note that the Eulerian fields can be defined in terms of functionals of $\mathbf{f}$ and its derivatives: 
\begin{equation}
\mathbf{x}:= \mathbf{f}(\mathbf{X},t) \,,\,
\mathbf{v}:= \dot{\mathbf{f}}(\mathbf{X},t) \, ,\, 
\mathbf{b}:= \ddot{\mathbf{f}}(\mathbf{X},t) \,,\,
\varrho(\mathbf{X},t)=\frac{\varrho_{\mathrm{ini}}}{J(\mathbf{X},t)} \,,\, 
\boldsymbol{\omega} = \frac{\boldsymbol{\omega}_{\mathrm{ini}}\cdot \boldsymbol{\nabla} \mathbf{f}}{J(\mathbf{X},t)}\,, \,etc. ,
\label{identi}
\end{equation}
where $etc.$ means any functional definition of other fields, written in terms of $\mathbf{f}$ and its derivatives, and where 
$\varrho_{\mathrm{ini}} (\mathbf{X})$ and $\boldsymbol{\omega}_{\mathrm{ini}}(\mathbf{X})$ stand for the initial values of the density and vorticity fields; an overdot denotes the partial time-derivative, and $\boldsymbol{\nabla}_{\bf 0}$ the nabla operator with respect to the Lagrangian coordinates.
Contrary to the Eulerian description, a single variable $\mathbf{f}$ describes the gravitational dynamics.
In the Lagrangian approach, the Eulerian position
$\mathbf{x}=\mathbf{f}(\mathbf{X},t)$ is no longer an independent
variable; the independent variables are now $(\mathbf{X},t)$.
According to the equivalence of inertial and gravitational masses, we
can express the field strength in terms of the acceleration
$\mathbf{g}=\mathbf{b}=\ddot{\mathbf{f}}$. The Eulerian evolution equations \eqref{ENS1} and \eqref{ENS2} reduce to definitions, but we have to transform the Eulerian spatial derivatives in the field equations \eqref{ENSelectric}:
\begin{eqnarray}
\label{LNS1}
\frac{1}{2} {\epsilon}_{\mathbf{abc}} {\epsilon}^{ikl}\ddot{f}^{\mathbf{a}}_{\;\,\vert i}  \, f^\mathbf{b}_{\;\,\vert k} \, f^\mathbf{c}_{\;\,\vert l} = \Lambda J - 4 \pi G \varrho_{\mathrm{ini}}\quad,\quad \delta_{\mathbf{ab}} \ddot{f}^{\mathbf{a}}_{\;\,\vert [i}\,  {f}^{\mathbf{b}}_{\;\,\vert j]} = 0\;.
\end{eqnarray}
This closed system of equations, the {\it  Lagrange-Newton system} \cite{buchertgoetz,buchert:lagrange,ehlersbuchert:lagrange}, appears to depend on the gradient of $\mathbf{f}$, but not on $\mathbf{f}$ itself. Since the constraints were transformed to evolution equations, the elliptic character of the Newtonian equations in Eulerian form seems to have got lost, but it re-surfaces in the construction of the initial data for the solutions of the Lagrange-Newton system.

We can express the {\it  Lagrange-Newton system} in terms of properties of the {\it  Newtonian tidal tensor} $E_{ij}$,  
\begin{equation}\label{eq:def_Eij}
E^i_{\;\; j} \equiv g^i_{\;\, , j} - \frac{1}{3}\delta^i_{\;j}g^k_{\; , k} 
=\frac{1}{2} \epsilon_{\mathbf{abc}}\epsilon^{ikl}\ddot{f}^\mathbf{a}_{\;\,\vert j} f^\mathbf{b}_{\;\,\vert k}  f^\mathbf{c}_{\;\,\vert l} + \frac{1}{3} \big( 4 \pi G \varrho_{\mathrm{ini}} - \Lambda J \big) \delta^i_{\;j} \; ,
\end{equation}
where we have used $\mathbf{g} = g^i\partial_i$ and its definition in terms of $\bf f$, together with inserting the first of the field equations (\ref{generalnewton}) into the divergence of the field strength. In terms of this form of the tidal tensor, written as a functional of the gradient of $\mathbf{f}$,
we can express the governing system through conditions on the tidal field:
\begin{equation}
\label{tidalformulationLNS}
   E_{[ij]} = 0 \qquad \textrm{and} \qquad E^k_{\;\; k} = 0 \; ,
\end{equation}
furnishing the four Lagrangian evolution equations for the three components of the trajectory field. 

In General Relativity we can derive the corresponding equations within a foliation of space-time into flow-orthogonal hypersurfaces of constant synchronous time for the matter model {\it irrotational dust}. 
Splitting the Einstein equations according to the $3+1$ formalism, the (local) Lagrangian coordinates can be used as Gaussian normal coordinates. Expressed in the local exact basis $\{\mathbf{d} X^i\}$, which spans the local cotangent spaces on the manifold, the dynamics of the fluid is no longer described by the gradient of the Newtonian trajectory function $\mathbf{d} f^\mathbf{a} = {f}^\mathbf{a}_{\ \vert i}\mathbf{d} X^i$, but by its  relativistic counterpart:  the non-integrable Cartan coframes $\boldsymbol{\eta}^\mathbf{a}= \eta^\mathbf{a}_{\ i} \mathbf{d} X^i$. The $4$-dimensional metric can be decomposed according to
\begin{equation}
\label{metric}
\mathbf{g}^{4} =  \mathbf{d}t \otimes \mathbf{d}t \;-\lambda \mathbf{g}^{3}
\quad \mbox{where}\quad \mathbf{g}^{3} = \delta_{\mathbf{ab}} \boldsymbol{\eta}^\mathbf{a} \otimes  \boldsymbol{\eta}^\mathbf{b}\:\:\Rightarrow g_{ij} = \delta_{\mathbf{ab}} \eta^\mathbf{a}_{\ i}\eta^\mathbf{b}_{\ j}\:.
\end{equation}
Inserting the metric expression \eqref{metric} into the Einstein field
equations, we  obtain a set of equations that contain the following
subset (that we may call the electric part of the
Lagrange-Einstein system, because this set is related to properties of
the spatially projected electric part of the Weyl tensor---as the
above Lagrange-Newton system is related to properties of the tidal
tensor, as we have shown) \cite{buchert:focus,rza1}):
\begin{equation}
\label{LESelectric}
\epsilon_{\mathbf{abc}}\epsilon^{ik\ell} \ddot{\eta}^\mathbf{a}_{\ i} \eta^\mathbf{b}_{\ k} \eta^\mathbf{c}_{\ \ell}   = \Lambda J - 4 \pi G {\varrho}_{\mathrm{ini}}
\quad,\quad \delta_{\mathbf{ab}} \,\ddot{\eta}^\mathbf{a}_{\ [i} \eta^\mathbf{b}_{\ j]} = 0 \;.
\end{equation}
This set of equations is formally similar to Eqs. \eqref{LNS1}, where $f^{i \rightarrow \mathbf{a}}_{\ \vert j}$ is replaced by ${\eta}^\mathbf{a}_{\ j}$.
However, in the former case it is a closed system, while in the latter case more equations for the nine components of the 
Cartan deformation are needed.

Thus, in a spatially diffeomorphism invariant language, the spatial Newtonian ``limit'' $\mathbf{d} f^\mathbf{a} \rightarrow \boldsymbol{\eta}^\mathbf{a}$
is rather a {\it restriction} of a general one-form to an exact one,
where exactness of the form implies the existence of an embedding vector
space. Although $\mathbf{d}\mathbf{d} f^\mathbf{a} = \mathbf{0}$, compared to $\mathbf{d}\boldsymbol{\eta}^\mathbf{a} \ne \mathbf{0}$ (leading to the Cartan connection one-form and, with another exterior derivative, to the curvature two-form in the 
Cartan structure equations), the Newtonian connection in the Lagrangian coordinate representation is nonzero \cite[Sec. III.A.4]{rza1},
\begin{equation}
{}^{N}\Gamma^i_{\ k\ell} = f^\mathbf{a}_{\ |k\ell} h_{,\mathbf{a}}^{\ \; i}
= \frac{1}{2 J( f^\mathbf{a}_{\ \vert j}) } \epsilon_{\mathbf{abc}}\epsilon^{imn} f^\mathbf{a}_{\ \vert k\ell} f^\mathbf{b}_{\ \vert m} f^\mathbf{c}_{\ \vert n} \;\ne 0\;\;,
\end{equation} 
where we expressed the matrix inverse to $f^\mathbf{a}_{\ \vert i}$, $h_{,\mathbf{a}}^{\ \; i} := \partial h^i / \partial x^\mathbf{a}$ (with $\mathbf{h} = \mathbf{f}^{-1}(\mathbf{x},t)$), in terms of the Jacobi matrix $f^\mathbf{a}_{\ \vert i}$ itself. The non-vanishing connection reflects the curvilinear (rotational) nature of the Lagrangian frame. But, the Newtonian connection is {\it integrable}. J\"urgen points out that \cite[p. 2]{GO} ``The decomposition of the gravitational connnection [...] into a flat gravitational connection and a vectorial gravitational force field is in fact locally always possible", but ``only unique under the certain condition---i.e. an additional global restriction, which is in general not valid for cosmological models---of an asymptotically flat gravitational connection."
He stresses here that this restricting condition is in general not possible in cosmological models (a remark which is echoed in the exact treatment of Newtonian cosmologies in \cite{buchertehlers}).
Further below he says \cite[p.~9]{GO}: ``It seems that there is no local condition [...] such that Newton's theory is a result of the restriction $\lambda =0$ in Einstein's theory with $\lambda > 0$."

Also note that Lagrangian observers are at rest, so that they do not ``see'' the light cone. The usual limiting process only applies to the $4-$dimensional coframes, where the restriction to exact forms represents the Einstein field equations in Minkowski space-time. This leaves the opening-up of the light cones as a separate limiting process in the above $3+1$ formulation of the Newtonian limit. We conclude that, even if we consider a Lorentz-covariant form of Newton's theory, the ``limit'' is degenerate in the sense that we have to impose \textit{local} integrability of the spatial deformation with the consequence that the limiting theory is embedded into a vector space, together with corresponding \textit{global} issues of non-uniqueness, unless boundary or fall-off conditions are specified.

\section{What is Newtonian theory?}

We propose a gedankenexperiment with the following troubling
remark. Looking at the Lagrange-Newton system that depends on the
gradient ${\mathbf d}{\bf f}$ in Eqs. \eqref{LNS1}, any realization (in terms
of a numerical realization) would actually make the spatial deformation non-integrable.\footnote{That is, the impossibility of numerically prescribing the components of the matrix ($f^\mathbf{a}_{\ \vert i}$) such that they exactly derive from vector components. The Lagrange-Newton metric components in the Lagrangian basis, ${}^N g_{ij} = \delta_{\mathbf{ab}}f^\mathbf{a}_{\ \vert i}f^\mathbf{b}_{\ \vert j}$, can only then be transformed via the inverse mapping ${\bf f}^{-1}$ to the Euclidean metric components ${}^N g_{ij} \equiv \delta_{ij}$ in the Eulerian basis.}
Hence, in practice, we would realize a non-exact one-form and
consequently we would create a non-integrable spatial  connection and
spatial curvature. In fact, we would look at the electric part of
Einstein's equations \eqref{LESelectric}.\footnote{This insight can be
  exploited to construct relativistic solutions from Newtonian ones
  \cite{rza3}.} Both on the level of equations and the form of
the solutions, the results would coincide and the difference 
would lie only in the Cauchy problem, i.e. the comparison of admissible initial data in General Relativity and those initial data that are admissible in the elliptic Newtonian initial value problem \cite{Lottermoser}. Einstein's theory delivers, together with Cartan's structure equations, the complementary set of equations needed to fully determine the Cartan deformation, hence the metric.

We can reformulate the above in another context. 
We look at a further common sense expression. The Newtonian theory has
a vanishing spatially projected magnetic part, $H_{ij}$, of the Weyl
tensor. From the geometrical perspective this is true and we can show
this also in Frame theory \cite{ehlersbuchert:weyl}. However,
J\"urgen notes \cite{GO} that the local equation of motion follows from the field equations in Einstein's theory, but for $\lambda=0$, ``the  structures and  laws  are  independent from each other in the limiting theory NG \cite[p.~16]{GO}''.
We will see that there is a non-vanishing term corresponding to the (vanishing) geometrical part $H_{ij}$ in the equations of motion that, however, decouples from the geometry in the limit. 
To see this we recall that the Einstein equations for an irrotational
dust  model can be cast into a system of evolution equations and
propagating constraints where the former reduce to a coupled system of
ordinary differential equations for the restmass density $\varrho$,
the rate of expansion $\Theta$, the rate of shear $\sigma^2 : = (\sigma^i_{\; j}\sigma^j_{\; i})/2$, and the trace of the spatially projected electric part of the Weyl tensor $E^2 : = (E^i_{\; j}E^j_{\; i})/2$ \cite{ehlers:fluids,EllisvanElst} by setting the corresponding magnetic part to zero.
These so-called \textit{silent universe models} (see \cite{henk:silent} and references to earlier work therein) are often used in cosmology. 
Now, the Newtonian theory also allows us to write the same set of evolution equations (where $2 E^2$ is the tidal tensor multiplied by itself),
but it is \textit{not} ``silent'', although the geometrical complement, the spatially projected magnetic part of the Weyl tensor, vanishes in the Newtonian limit. 
Instead, there is a term that links the system of evolution equations to the (non-local) constraints, the absence of which would again restrict the class of solutions drastically, see \cite{kofmanpogosyan,ehlersbuchert:weyl}.  

These properties of Newtonian equations on the one hand illustrate what J\"urgen means by the fact that the limiting procedure provides equations of motion that are decoupled from the geometry, or as already quoted above that ``the {\it formal} simplification is connected to a {\it conceptual} complication \cite[p.~16]{GO}'', but on the other hand they point to the reason for the decoupling of the evolution equations from the geometry: this is furnished by the restriction to integrability of Cartan deformations and connections, which is purported by the word 
``degenerate limit'', and which is strictly speaking independent of the limiting process of sending $\lambda$ to zero.

While enjoying J\"urgen's paper, the reader may especially look at J\"urgen's list of questions in \cite[Section 3]{GO} as a guide for reading.
J\"urgen answers aspects of these questions, but a number of issues remain open until today. We therefore contemplate some further aspects of J\"urgen's questions in what follows.

\section{J\"urgen's fifth question}\label{sec:5th}

J\"urgen's fifth question \cite[p.~12]{GO} deals with possible approximation formulae for one-parametric families of metrics $g_{ab}(\lambda)$
in Einstein's theory of gravitation following from the existence of a Newtonian limit of such families of metrics. 
In particular, the question asked is: 
  \begin{quote}
{\it (5a) ``Do approximation formulae, e.g., for asymptotic representations, for the fields $g_{ab}(\lambda), \hdots$ for $\lambda\rightarrow 0$,  follow from the existence of a Newtonian limit of a family of Einsteinian solutions?''}
\end{quote} 
 to which he added in his personal printout of the article, encoded in handwritten notes, what is the 
 \begin{quote}
{\it (5b) ``differentiability for $\lambda$ at $\lambda=0$?''}
\end{quote}
Soon after publication of J\"urgen's article, Jeffrey Winicour
presented a series of articles \cite{newt1,newt2,quad} answering a
modified version of J\"urgen's fifth question, and he also makes
statements on the differentiability with respect to
$\lambda$ at $\lambda=0$.
Winicour's motivation is the  numerical application of such approximation formulae for the characteristic formulation of General Relativity and especially its relation to the Bondi-Sachs formulation \cite{Bondi,Sachs} (see e.g. \cite{BSscolar} and \cite{JeffLRR} for review) of the latter. 
 His work was commented by his contemporary colleague Bernard Schutz as \cite{Schutz} -- ``...The null cone approach to the Newtonian limit described by Winicour is one of the most original ideas to have emerged in this subject recently...''.

In the Bondi-Sachs formalism, Einstein's space-time of General Relativity is foliated by a family of (outgoing) null hypersurfaces. 
This has the advantage that, if the coordinate along the generators of the null hypersurfaces is suitably compactified,  the Penrose compactification scheme of space-time \cite{Penrose} can be performed  with a suitable conformal factor \cite{tam}. 
In turn, null infinity can be attached to the physical space-time and gravitational waves can be read off at null infinity, where they are rigorously defined. 
The gravitational waves are encoded in the shear $\sigma$ of the outgoing null hypersurfaces.  
This shear is the crucial ingredient for the initial data to solve Einstein's equations for a Bondi-Sachs metric.
By considering the Einstein-fluid equations, Winicour investigates J\"urgen's question  by considering the two  questions \cite{newt1}: 
\begin{quote}
{\it (1) What is the appropriate gravitational  data such that the limit $\lambda\rightarrow 0$ gives the Newtonian gravitational structure for this fluid?}
\end{quote}
Provided the  $\varrho_N$,  $v^i_N$ and $\sigma_N$ are the data for the Einstein-fluid equation of a Bondi-Sachs metric $g_{ab}(\lambda)$  on an interval $[u_0,u_1]$ of retarded times $u\in[u_0,u_1]$, and $\varrho_N$, $v^i_N$ and $\sigma_N$ are the density, the three velocity $v^i_N$ and the gravitational shear of the corresponding Newtonian limit of $g_{ab}(\lambda)$ at $\lambda=0$, then \cite{newt1}:
\begin{quote}
{\it (2) Do $\varrho_N$,  $v^i_N$ and $\sigma_N$ remain valid throughout this time interval, or does the $\lambda=0$ limit of the Einsteinian fluid evolve away from its Newtonian counterpart?}
\end{quote}
\begin{figure}[htbp]
\begin{center}
\begin{center}
  \includegraphics[width=0.6\textwidth,draft=false]{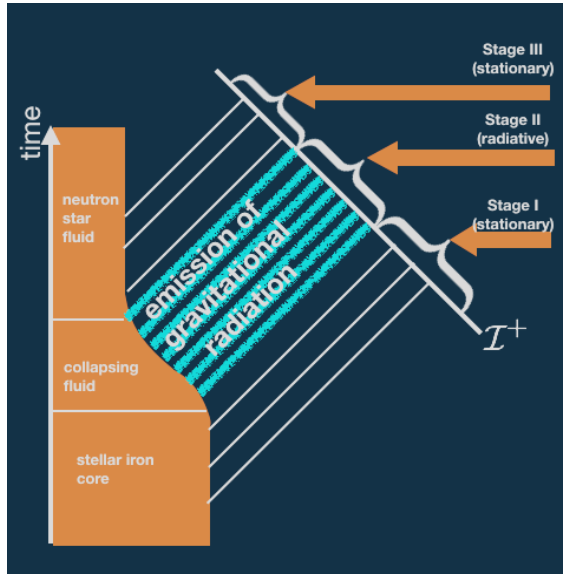}
\end{center}
\caption{\label{Fig:model} Example for a stationary-to-radiative-to-stationary three-stage model of a stellar core-collapse described in the text. Null infinity is the surface $\mathcal{I}^+$. The time axis measures the Newtonian time $t$ along the centre of mass of the stellar fluid.}
\end{center}
\end{figure}
An exemplary system with a Newtonian limit is a  core-collapse supernova scenario, where the initial system, say (I), is an degenerated iron core of an isolated star of mass\footnote{The solar mass $M_\odot$ is $M_\odot\approx2\times10^{30}$kg.}  $M<8M_\odot$ (see e.g., \cite[for an astrophysical review]{Muell,review2}\cite[in the context of the Bondi-Sachs formalism]{Siebel} and \cite[where a star collapses to a black hole via formation of an intermediate 
(proto-)neutron star (see Figure~\ref{Fig:model})] {cerda}\footnote{In this
case the quasi-Newtonian initial data of \cite{newt1,newt2} can also be employed. However, the final space-time will contain a Kerr black hole whose Newtonian limit in Frame theory is a point mass \cite[p.~9]{GO} and \cite{ehlers:examples}. As such Figure~\ref{Fig:model} is only valid up until the formation of the (quasi-Newtonian) 
proto-neutron star.}.
The state (I) is `clearly' Newtonian as can be seen by calculating the relevant compactness $C = GM / ( c^2R )$ of the object, i.e., $C\lesssim 10^{-6}$.  
The iron-core collapses due to the lack of radiation pressure, caused by the stopping of the nuclear fusion in its centre, to balance the gravity of the fluid.
During the collapse phase,  (II), of the fluid, gravitational waves are in general emitted from the region where the fluid resides.
This is  because the collapsing fluid behaves to lowest order as a
time-varying quadrupole since the initial model (I) is in general
rotating.\footnote{Only in (the astrophysically unlikely case of)  a
  spherically symmetric scenario, are no gravitational waves emitted.} 
The final remnant of stage (III) after the  collapse is a proto-neutron star of mass $M\lesssim1.4 M_\odot$, which is a weak relativistic system where  $C\sim 10^{-1}$.  
In this example, the physically investigated system gives rise to a space-time that does not have ``holes'' in the sense required by J\"urgen \cite[see Sec. 3]{GO}. 
That is, space-time has a compact space containing the stellar fluid and an empty region commencing at the boundary of the fluid. 

Since all of the mass is confined to a compact space, it is possible to single out a centre-of-mass world line of the fluid. 
It is thus natural to choose freely falling (Fermi) observers \cite{Manasse} along this world line, and choose the outgoing null cones emanating from this world line to define a one-parametric family of Bondi-Sachs metrics  $g_{ab}(\lambda)$ (such a construction is demonstrated in the axisymmetric vacuum case in \cite{vertex}). 
Studying the time evolution of $g_{ab}(\lambda)$ from stage (I)  via the radiative period (II) to the final stage (III) has mainly  been done numerically, because on the one hand the equation of state for the fluid matter is in general provided in tabular form, while on the other hand the underlying Einstein-fluid equations are highly nonlinear. 
However, the numerical solution provides an answer to J\"urgen's question stated above if $\lambda=0$ is chosen for the numerical solution of the one-parametric family of metrics $g_{ab}(\lambda)$. 

Despite the problems of nonlinearity and the issue with the equation of state, Winicour  is able to make qualitative statements on the evolution of the initial data.
But, before a numerical study can be made, the initial data at stage (I) for the Bondi-Sachs metric at some retarded time $u=0$ must be provided. 
This requires the `correct' determination of the gravitational shear $\sigma$ 
for the Newtonian system in the (stationary) stage (I).
Winicour's two main articles \cite{newt1} and \cite{newt2} deal  with this question. 
For its answer, the following assumptions on the Newtonian initial model at stage (I) are made: (i) the fluid is a perfect fluid with 
matter density  $\varrho$ and an equation of state $p(\varrho,\lambda)$, (ii) $w^i$, the three-dimensional velocity field of the matter, 
(iii) $\Phi$, there is a Newtonian gravitational (connection) potential, (iv)
the matter is coupled with the connection potential via the Poisson
equation, and (v) the (non-relativistic) Euler equations. 

\section{Opening up light cones---part II}

Before discussing the $\lambda$-families of Bondi-Sachs metrics $g_{ab}(\lambda)$, it is worth looking at how a 
one-parametric family $\eta_{ab}(\lambda, x^c)$ of the Minkowski metrics in standard Cartesian coordinates $x^a=(t,x^i)$,
\begin{equation}
\label{eq:mink}
\eta_{ab}(\lambda, x^a) {\mathbf d} x^a \otimes {\mathbf d}x^b = \mathbf{d} t\otimes\mathbf{d} t -\lambda  \delta_{ij} \mathbf{d} x^i \otimes\mathbf{d} x^j\;\;,
\end{equation}
gives rise to a flat space version of a family of Bondi-Sachs metrics $\eta_{ab}(\lambda, y^a)$ with Bondi-Sachs coordinates $y^a$. 
We denote with $t_{a} = t_{,a}$ the covector of the surfaces of constant time $t$.
The limit of \eqref{eq:mink} for $\lambda=0$ provides an empty Newtonian space-time structure with 
$$g_{ab} = t_at_b\;\;,\;\; \lambda g^{ab} = -\delta^a_{\phantom{a}i}\delta^b_{\phantom{a}j}\delta^{ij}\;\;,$$
where the second of the two equations defines J\"urgen's spatial metric \cite[p.~4]{GO}, 
$${h^{ab}:=- \lim_{\lambda\rightarrow0}(\lambda g^ {ab})}\;.$$
Introduction of  $r^2 = \delta_{ij}x^i x^i$, allows us to define the scalar field 
$$u = t - \lambda^{1/2} r(x^i) = t-\lambda^{1/2}\big[(x^1)^2+(x^2)^2+(x^3)^2\big]^{1/2}\;,$$
  which corresponds to the special relativistic retarded time\footnote{In units where the speed of light equals unity.} $u$, if $\lambda=1$, and to the Newtonian `absolute time', if $\lambda=0$.
  The gradient to the surfaces $u=const$ is
   \begin{equation}
	\label{eq:tangent}
k_{a} = u_{,a}= (1, -\lambda^{1/2}r_{i})\;\;,\;\;r_i:=r_{,i}\;\;.
\end{equation}
   The squared norm of $k_a$ vanishes with respect to $\eta_{ab}(\lambda, x^a)$, thus hypersurfaces $u=const$ are null hypersurfaces.
   In fact, they are light cones with respect to the (straight) world line $c(t)$ passing through all of the points where $x^i=0$ in the hyperplanes $t=const$.
  The function $r$ measures the distance between the vertex of a given cone $u=0$ and points on this respective cone. 
  The three-dimensional vector $r_i$ measures the direction cosines with respect to a parallel propagated frame along 
  $c(t)$.
  As for  the relation $x^i = rr^i$, it is useful to parameterize $r^i = r^i(y^A)$ by two angular coordinates $y^A$ that are constant along the generators of the cone. 
   Then $y^A$ measure the direction angles of light rays emanating from the vertex along the light cone $u=0$. 
   The `propagation velocity' of light along these rays is $\lambda^{-1/2}$. 
   Now, consider one arbitrary light cone $u=0$ at proper time $t=0$ on $c(t)$. Letting $\lambda\rightarrow 0$, we observe in the behaviour of the tangent vector \eqref{eq:tangent} the  `opening up' of the light cones mentioned in  J\"urgen's  introduction. 
   In this limit, $k_a\rightarrow t_a$, and $r$ measures the space-like distance in the hyperplane $t=0$ with respect to the point $x^i=0$, while $y^A$ measure spherical angles with respect to the origin $x^i=0$. 
   
   A one-dimensional family of flat space Bondi-Sachs metrics $\eta_{ab}(\lambda, y^A)$ along a world line $c(t)$ is found by a simple coordinate transformation of \eqref{eq:mink} 
   to null polar coordinates $y^a = (u, y^1=r, y^A)$, so that 
  \begin{equation}
	\label{eq:minkBS}
\eta_{ab}(\lambda, y^a) {\mathbf d}y^a \otimes {\mathbf d}y^b = {\mathbf d}u\otimes {\mathbf d}u  + 2\lambda^{1/2} {\mathbf d}u\otimes {\mathbf d}r -\lambda r^2 q_{AB}(y^A) {\mathbf d}y^A\otimes {\mathbf d}y^B \;,
\end{equation}
  where $q_{AB}(y^C) = r^i_{,A}r^j_{,B}\delta_{ij}$ is the unit round sphere metric.
  We take $y^A$ to be the standard spherical angles $y^A = (\theta, \phi)$ such that  $q_{AB} = \mathrm{diag}(1,\sin^2\theta)$.  The metric \eqref{eq:minkBS} also shows in the limit $\lambda\rightarrow 0$ that 
  \begin{equation}
\eta_{ab}(\lambda, y^a)|_{\lambda=0} = u_{,a}u_{,b} = t_at_b\;\;,\;\;
 h^ {ab} = -\delta^a_r\delta^ b_r - \frac{1}{r^2}\delta^ a_A\delta^b_B q^ {AB}(y^ C)\;,
\end{equation}
where $q^{AB}$ is the inverse metric of the unit sphere $q^ {AB} = \mathrm{diag}(1, \sin^{-2}\theta)$, and we adopted 
J\"urgen's notation for the inverse spatial metric $h^{ab}$. 

\section{One-parametric families of Bondi-Sachs metrics $g_{ab}(\lambda, y^c)$ }

For the investigation of the Newtonian limit of an Einstein-fluid system, the above construction needs to be done at the centre-of-mass world line of the initial model at stage (I).
   The regularity requirements of the metric along the centre-of-mass world line imposed by a freely falling observer introduces constraints on the metric at the vertices of the null cones that have to be maintained throughout the evolution of the system from stage (I) to stage (III).
   As such, the boundary conditions are fixed at the origin of the system, which is contrary to fixing the boundary conditions at infinity as proposed by J\"urgen \cite[see Theorem 1 and the  conjecture in Sec.~3]{GO}. 
   The reason for the former is that the Einstein equations in their Bondi-Sachs formulation are integrated along the null rays emanating from the vertex, and the vertices of the null cones must follow a regular world line.
   Moreover, setting up a Bondi evolution at null infinity is ill-defined as it requires data, the news function $N:=\frac{1}{2}\sigma_{,u}$,  in the future because $N$ is a retarded time derivative. 
   
A one-parametric family of Bondi-Sachs metrics $g_{ab}(\lambda, y^a)$  is given by \cite{newt1,newt2}:
\begin{eqnarray}
g_{ab}(\lambda, y^a) {\mathbf d}y^a\otimes {\mathbf d}y^b&=&
	\Big (\frac{V}{r}e^{2\lambda\beta} - \lambda^2 r^2 h^{AB}U_AU_B \Big) {\mathbf d}u\otimes {\mathbf d}u + 2\lambda^{1/2} e^{2\lambda \beta} {\mathbf d}r \otimes {\mathbf d}u \nonumber 
	\\&&
 +2 \lambda^{3/2} r^2U_A {\mathbf d}x^A \otimes {\mathbf d}u
	 - \lambda r^2 h_{AB} {\mathbf d}y^A \otimes {\mathbf d}y^B \;,
	\label{eq:metricBS}
	\end{eqnarray} 
	where the metric functions $V$ and $h_{AB}$ are represented by 
	\begin{equation}
	\label{mf}
	V = r + \lambda W\;\;,\;\;
	h_{AB} = q_{AB} + \lambda \gamma_{AB}\;,
	\end{equation}
	so that, if $W=U_A=\gamma_{AB} = 0$, the Minkowski metric \eqref{eq:minkBS} is obtained.\footnote{Comparison of \eqref{eq:metricBS} with the corresponding metrics in Winicour's pioneering articles \cite{newt1,newt2,quad} shows that we have set $\lambda = \lambda^2_{W}$, to be in tune with J\"urgen's article where the causality constant $\lambda$ is represented by $\lambda = 1/c^2$.} 
	The $\lambda$-factors in \eqref{eq:metricBS} are chosen such that the Einstein equations ensure a smooth limit for the fields $\beta, W, U_A$ and $\gamma_{AB}$  at $\lambda=0$. This partially answers J\"urgen's (handwritten) question on the differentiability for $\lambda=0$. 
        The conformal metric $h_{AB}$ has the determinant of the unit sphere $\det(h_{AB}) = \det(q_{AB}) = \sin^2\theta$. 
        Setting $\lambda=1$ in \eqref{eq:metricBS} yields the standard form of the Bondi-Sachs metric \cite{Bondi,Sachs,BSscolar}.
        The symmetric $2-$tensor $\gamma_{AB}$ is transverse-traceless and, due to the choice of coordinates, it gives rise to  the shear tensor $c_{AB}$ of the outgoing null cones with tangent vector $k^a$,
       \begin{equation}
\label{eq:sheartensor}
 c_{AB} := \gamma_{AB,r} = \lambda^{-1}h_{AB,r}\;\;.
\end{equation}
        The gravitational shear tensor incorporates the $\oplus$ and $\otimes$ polarizations of the gravitational waves (see, e.g., \cite{Thorne} for a relation to the `usual' post-Newtonian formulation). 
        Given a complex dyad $m^A$ defined via $$h_{AB}m^A\bar m^B=1\;\;;\;\;
        h_{AB}m^A m^B= 0\;\;;\;\;h_{AB} = 2m_{(A}\bar m_{B)}\;,$$ the two degrees of freedom of $c_{AB}$ are most conveniently expressed through a complex spin-weight-two   
        scalar field,
        \begin{equation}
	\label{eq:shearNP}
	\sigma = m^Am^B c_{AB}\;\;,
	\end{equation}
        because the decomposition of $c_{AB}$ in terms of $m_A$ is 
        \begin{equation}
\label{eq:shearBS}
c_{AB} = \sigma \bar m_A \bar m_B + \bar \sigma m_Am_B\;\;.
\end{equation}
        The variable $\sigma$ is one of the Newman-Penrose connection coefficients,
        and it is the free gravitational datum of the characteristic initial value problem. 
         Considering the Newtonian limit of the curvature, $\sigma$  determines part of  the Weyl tensor
        \begin{equation}
	\label{eq:Weyl}
	m^Am^B C_{rArB}\Big|_{\lambda=0} = -  \frac{1}{2}(r^2 \sigma)_{,r}\;\;.
	\end{equation}
	As this part of the Weyl tensor corresponds to the
        Newman-Penrose scalar $\Psi_0$, which represents the incoming
        radiation of a system, the `correct' choice for the free data
        has an important bearing on the existence
        of a Newtonian limit given a one-parametric family of
        Bondi-Sachs metrics $g_{ab}(\lambda, y^a)$. This means that,
        although $\sigma$ is, in principle,  arbitrary---like, for
        example,  setting the initial shear to zero at some time
        $u=0$---this will introduce incoming radiation to the system \cite{newt1} (see e.g. \cite{BKS,BoostedKerr} for    
        details on ingoing radiation in Schwarzschild/Kerr space-times).
   
\section{The Newtonian limit of a family of null cones}
   
   Due to the implications of the twice contracted Bianchi identities and due to the requirement of the regularity conditions imposed at the vertices of the null cones, only six of the ten Einstein equations need to be considered for the 
   Bondi-Sachs metric \eqref{eq:metricBS} (see e.g. \cite{BSscolar} for demonstration). Four of these six equations are hypersurface equations not containing any time-derivatives, and the remaining two equations describe the retarded 
   time-evolution of $\sigma$. Only the general structure of these equations is required here, not their particular form.
   
   \subsection{Consequences of the hypersurface equations}
   Two of the four hypersurface equations are second-order differential equations, and it is illustrative to rewrite them in terms of four first-order equations. Moreover, we also list the definition of the shear as one of the hypersurface equations.   The final hypersurface equations to be considered are:
  \begin{subequations}\label{eq:hierarchy}
   \begin{eqnarray}
   h_{AB,r}&=& F^{(0)}_{AB}( \sigma) \;\;;\label{eq:hAB}\\
   \beta_{,r} & = & F^{(1)}( \sigma, h_{AB}, T_{rr})\;\;;\label{eq:beta} \\
   Q_{A,r} & = & F^{(2)}_A( \sigma, h_{AB},\beta, T_{rA})\;\;; \label{eq:Qa}\\
   h_{AB}U^B_{,r} & = &  \frac{1}{r^4} e^{-2\lambda \beta}Q_{A}\;\;;\label{eq:Ua}\\
   W_{,1} & = &  F^{(3)}( \sigma, h_{AB},\beta,Q_A, U^A , T_{ru})\;\;,\label{eq:W}
\end{eqnarray}
\end{subequations}
 where $U^A = h^{AB}U_B$, and the placeholders $F^{(\bullet)}_{\bullet}$ are in general nonlinear functions of the arguments. 
 The right-hand side of \eqref{eq:hAB} follows from the definitions \eqref{eq:sheartensor}-\eqref{eq:shearBS}.
 The four equations \eqref{eq:Qa} and \eqref{eq:Ua} are the result of casting the respective second-order differential equations of the corresponding Einstein equations into first-order equations. 
 The components $T_{ra}$ are the components of the energy-momentum tensor of a perfect fluid with an isotropic pressure $p$ of the form:
\begin{equation}
T_{ab} = (\varrho +\lambda p)w_aw_b -\lambda p g_{ab}\;\;, w_a = t_a + \lambda v_a\;\;,
\end{equation}
where $v_a$ is the $4-$velocity of the fluid with the Newtonian limit ${v^a = (1,V^r, V^A)}$ in polar coordinates for $\lambda=0$.
Equations \eqref{eq:hierarchy} show that, if the values of the $h_{AB}$, $\beta$, $Q_A$, $U^A$ and $W$  are known at the vertices $r=0$,\footnote{These values follow from the requirement of having a freely falling origin at $r=0$.} then this provides knowledge of 
$\sigma$ and $T_{ra} (\varrho, v_a)$ everywhere on a null cone $u=u_0$;
the equations \eqref{eq:hierarchy} can then be solved in a hierarchical manner. 
For example, given $\sigma$ on a cone $u_0$ and initial values for $h_{AB}$ at the vertex,  \eqref{eq:hAB} can be solved for $h_{AB}$ on the entire cone $u_0$; then, with the so-obtained $h_{AB}$, $\beta$ can be found on the cone $u_0$ using \eqref{eq:beta}, provided we have knowledge of $T_{rr}$ on all of $u_0$ and an initial value for $\beta$.
This hierarchical scheme proceeds until $W$ is found after integration of \eqref{eq:W}.
In particular, we note that $W$ is completely determined by the shear $\sigma$, the fluid data $\varrho$ and $v_a$, that is 
$$W = W(\sigma, \varrho, V_r, V_A)\;\;.$$ 
Turning the argument around, this also implies that any restriction on $W$ must necessarily restrict the gravitational shear $\sigma$, meaning
$$\sigma = \sigma(W, \varrho, V_r, V_A)\;\;,$$
provided the hierarchy \eqref{eq:hierarchy} can be inverted in a given way. 
Indeed, considering Eqs. \eqref{eq:hierarchy} for the limit $\lambda=0$, Winicour has shown \cite{newt1,newt2} that there exists a scalar field $\Phi^*$ given by 
\begin{equation}
\Phi^* = \frac{W}{2r}\Big|_{\lambda=0} + \beta|_{\lambda=0}\;,
\end{equation}
that is $O(r)$ at the vertices $r=0$, and which allows for such an inversion. 
On top of that,  this scalar field has  the desired decomposition  of the connection \cite[Eq. (39)]{GO},
 \begin{equation}
\Gamma^a_{bc} = \Big(\stackrel{o}{\Gamma}\!^a_{cd} - \lambda t_bt_c g^{ae}\Phi^*_{,e}\Big)\Big|_{\lambda=0}\;,
\end{equation}
into the connection of a pure field of inertia $\stackrel{o}{\Gamma}\!^a_{bc}$ (in spherical coordinates $r$, $\theta$ and $\phi$) and a space-like vectorial gravitational field  $g^{ae}\Phi^*_{,e}$ in the Newtonian limit $\lambda=0$, which also obeys the Poisson equation in the limit $\lambda=0$. 
Denoting with $Y_{lm}(y^A)$ the standard spherical harmonics, $\Phi^*$ differs from the usual Newtonian potential $\Phi$ by  monopole-dipole terms:
\begin{equation}
\label{eq:modPhi}
\Phi^* =  \Phi + aY_{00}(y^A) + r [a_{-1}Y_{1-1}(y^A)+a_{0}Y_{10}(y^A)+a_{1}Y_{11}(y^A)]\;,
\end{equation}
which vanish at infinity through a solution of the Laplace equations. 
Here, $\Phi$ is a solution of the Poisson equation of the Newtonian initial model. 
By a rather tedious inversion of  the hierarchy \eqref{eq:hierarchy}, Winicour has shown that the modified Newtonian potential relates to a shear-like term, 
\begin{equation}
\label{eq:Phi*}
q^Aq^B \eth_A\eth_B\Phi^* = -\frac{1}{2} (r^2 \sigma)_{,r}\Big|_{\lambda=0}\;,
\end{equation}
where ${q^A = 2^{-1/2}(1, i\sin^{-1}\theta)}$ and  $\eth_A$ is the covariant derivative of the unit round sphere metric. 

Relation \eqref{eq:Phi*} is in principle the answer to Winicour's question (1) and it gives rise to a `simple' algorithm to determine the initial data, i.e., $\sigma$,  for a  quasi-Newtonian system at a null cone with a Newtonian limit for $\lambda=0$: first, the Poisson equation for the potential $\Phi$ is solved for the Newtonian system; 
second, the gravitational shear $\sigma$  is determined from \eqref{eq:modPhi} and \eqref{eq:Phi*}, where the boundary conditions are fixed at the origin.
Note that, as $\sigma$ is a complex field of spin-weight-two, it can always be written as
$\sigma = q^Aq^B\eth_A\eth_B \Sigma$,
where $\Sigma$ is a complex scalar field of spin-weight-zero. The real and imaginary parts of $\Sigma$ are called the
electric and magnetic parts of $\Sigma$, respectively. This nomenclature is adopted from the electric/magnetic decomposition of the Faraday tensor of the electromagnetic field, and corresponds to a curl/gradient decomposition of tensor fields on a sphere (see e.g. \cite{skypattern} for a related discussion with regard to gravitational waves). 

\subsection{Consequences of the evolution equations}

Considering the evolution equation of the Bondi-Sachs metric, which schematically  reads:
\begin{equation}
\label{eq:ev}
2(r^2\sigma)_{,u} = \lambda (r^2\sigma)_{,r}+ F^{(4)}(\lambda, \beta, h_{AB}, W, U^A, Q_A, \varrho, p, v_a)\;\;,
\end{equation}
Winicour shows that $\Sigma$ (and consequently also $\sigma$) must be
purely electric in order to provide initial data on the null cone that
are smooth in $\lambda$.\footnote{It is intriguing to remark at this
  stage that most physically relevant quantities are related to
  the electric (gradient) type of a tensor field rather than the
  magnetic (curl) type.  For example, in cosmological applications
  (e.g. \cite{ehlersbuchert:weyl} and Eq. (\ref{eq:def_Eij})) it is
  the electric part of the Weyl tensor driving most of the dynamics,
  the relevant initial data $\sigma$ on a null cone for an isolated system are of electric
  type and in the global asymptotic properties of gravitational
  radiation \cite{skypattern,supertranslation}, too,  only the electric type
  seems to play a fundamental role.}
To some extent this observation also gives a further answer to
J\"urgen's question (5b), indicating that the
differentiability in $\lambda$ is necessary to obtain a bona fide
Newtonian limit of his ``Frame Theory''. 

The answer to Winicour's second question (2) on the validity of the initial data with regard to its Newtonian limit during a retarded time-evolution requires consideration of the evolution equation \eqref{eq:ev} and the 
 local conservation law of energy-momentum, $\nabla_aT^a_{\phantom{a}b}=0$,  at leading order in $\lambda$. 
Energy-momentum conservation in the limit $\lambda=0$ ensures that $\varrho_N$ and $v^i_N$ remain valid throughout the time interval $[u_0, u_1]$.
To investigate the time-evolution of $\sigma_N$ for the evolution on the retarded  time interval $[u_0, u_1]$, all  fields ${\mathcal{F}\in\{\beta, h_{AB}, U^A, V, \varrho, p, v_a\}}$ are assumed to have a smooth $\lambda-$expansion of the type 
$\mathcal{F}(\lambda) = \sum_n \mathcal{F}_{(n)} \lambda^n$.
Inserting these expansions into the evolution equation \eqref{eq:ev} yields an iterative scheme, where the retarded time-derivative of  the shear  $\sigma_{(n)}$ is determined by the $\sigma_{(n-1)}$ coefficient. 
The $\sigma_{(1)}$ coefficient is unconstrained and may be used to fix the initial data on the cone $u=u_0$. 
The successive determination of the $\sigma_{(n)}$ from a Newtonian model $(\Phi, \varrho, p(\varrho), V_r, V_A)$ provides the gravitational initial data for the Bondi-Sachs system, preserving the Newtonian limit for $\lambda=0$. 
Proper determination of the $\sigma_{(n)}$'s  is crucial for ruling out, in J\"urgen's words, ``...incoming fields from the exterior. \cite[p.~9]{GO}''.
The algorithm to calculate $\sigma_{(n)}$ up to $n=2$ is given in \cite{newt2}, and an example for Newtonian dust at $\lambda=0$ can be found in \cite{dust}. 
 
We finally note an important observation regarding the
$\lambda$-expansion of $\mathcal{F}$: since all fields become
$\lambda$-dependent for $u>u_0$ during the evolution of the centre-of-mass world line of the Newtonian system, this system will in general move away from the geodesic along which the null cones of the Bondi-Sachs metric are constructed. 
This is because of the $\lambda$-dependence of the pressure, whose gradient introduces an acceleration towards fluid elements at the origin of the geodesic of the Newtonian theory at $\lambda=0$. 
 
\section{Closing remarks}\label{sec:close}
 
J\"urgen conjectured \cite[Sec.~3]{GO} that space-times that are asymptotically flat at space-like and null infinity have a Newtonian limit. 
This fixes the boundary conditions of the fields at infinity.
Asymptotic flatness at null infinity implies the peeling property \cite{Sachs} of the Weyl tensor $C_{abcd}$ in the limit $r\rightarrow\infty$. 
The ten independent components of $C_{abcd}$ are most conveniently expressed in terms of the five complex Newman-Penrose Weyl scalars $\Psi_n, n\in\{0,...4\}$ \cite{NP,NPscolar},
and the peeling property of asymptotically flat space-times is given by $\Psi_n(y^a) = \Psi^{(0)}_n(u, y^A)/r^{5-n} + O(r^{-6})$.
Winicour, however, fixed the boundary conditions of the fields at a geodesic inside a Newtonian matter distribution. 
This is necessary to obtain a well-defined characteristic initial value formulation. 
In turn, no conditions at null infinity can be fixed. 
Surprisingly, further investigations of Winicour and collaborators show, for an exemplary system with a Newtonian limit at $\lambda=0$, that the Weyl tensor does not need to have the peeling property. 
Especially due to the presence of a logarithmic term in $\sigma$ implied by such a Newtonian limit, the asymptotic behaviour of the Weyl tensor is spoiled \cite{quadPRL}, 
as it does not behave as $C_{abcd}=O(r^{-1})$, as implied by the peeling theorem, but it behaves as $O(r^{-1} \ln r)$. 
Thus, peeling cannot be recovered, but it is possible to establish a weaker version of the peeling theorem, where both the electric and the magnetic parts of the Weyl tensor are $O(r^{-1})$ in the limit $r\rightarrow\infty$ \cite{logScri}. 
Nevertheless, the famous Einstein quadrupole formula is unaffected by this logarithmic behaviour and has a Newtonian limit for $\lambda=0$ \cite{quad,quadPRL}. 
Although, the finding of logarithmic terms at null infinity came as a surprise when mapping a Newtonian system onto a null cone of a quasi-Newtonian relativistic system, logarithmic terms can also be found by considering the global asymptotics of the characteristic initial value formulation of General Relativity \cite{poly1,christ}.
Indeed, further investigations \cite{poly1} showed the existence of an entire `zoo' of polyhomogeneous space-times  (i.e., space-times where the Weyl tensors are $O(r^{-k}\ln^{m}r)$ with with $k,m\ge 1$), and whose mathematical meaning is not yet fully understood.
One might speculate that some of them relate to higher-order corrections to the null cone initial data when Newtonian models of a Newton-Cartan theory at $\lambda=0$ are mapped onto a null cone. 

J\"urgen  raises some doubts (``For Einstein's Theory many \textit{asymptotic flatness} conditions were proposed [...] but their suitability for non stationary space-times is still in doubt \cite[p.~9]{GO}'') about whether asymptotic flatness for isolated systems at null infinity is in fact a completed chapter.
Winicour's approximation model for a one-parametric  family of Bondi-Sachs metrics $g_{ab}(\lambda)$ with a  Newtonian system at $\lambda=0$ together with the zoo of polyhomogeneous space-times is proof that J\"urgen's doubt on asymptotically flat space-times is still standing.
In particular, J\"urgen's advice to not take for granted generally imposed assumptions (as noted in \cite{GO}
 ``...textbook and monographs suggest the existence of a well-understood [Newtonian] limit $(\rightarrow)$ ... \cite[p.~2]{GO}'') should be taken as a motto for any scientist, not only working in General Relativity. 

\section*{Biography: J\"urgen Ehlers}
A biography of J\"urgen Ehlers was published in \href{https://doi.org/10.1007/s10714-009-0841-7}{\textit{Gen. Relativ. Gravit.} \textbf{41}, 1899} (2009), in the form of the obituary reprinted from
the Biennial Report 2006/2007 of the Albert Einstein Institute.

\vspace{15pt}
\noindent{\bf Acknowledgements}
We wish to thank Malcolm MacCallum for inviting this editorial note and for constant assistance.
In particular, TM thanks Malcolm MacCallum for help on some subtle grammatical issues as well as `false friends'  appearing in a first draft of the translation of \cite{GO}. We thank Frank Steiner for valuable remarks.




\end{document}